\documentclass[aps,prl,twocolumn,groupedaddress,superscriptaddress]{revtex4-1}

\usepackage{amsmath}
\usepackage[pdftex]{graphicx}% Include figure files

\newcommand{\ud}[1]{#1^{\dag}}

\begin{document}

%\preprint{}

\title{Climbing the Jaynes--Cummings ladder by photon counting}


\author{Fabrice P.~Laussy}
\email[]{fabrice.laussy@gmail.com}
\affiliation{Walter Schottky Institut, TU M\"unchen, Am Coulombwall 4, D-85748 Garching, Germany}%
\author{Elena del Valle}
\affiliation{Physikdepartment, TU M\"unchen, James-Franck-Str. 1, D-85748 Garching, Germany}%
\author{Michael~Schrapp}
\affiliation{Walter Schottky Institut, TU M\"unchen, Am Coulombwall 4, D-85748 Garching, Germany}%
\author{Arne Laucht}
\affiliation{Walter Schottky Institut, TU M\"unchen, Am Coulombwall 4, D-85748 Garching, Germany}%
\author{Jonathan J.~Finley}
\affiliation{Walter Schottky Institut, TU M\"unchen, Am Coulombwall 4, D-85748 Garching, Germany}%
\homepage[]{http://laussy.org}
%\thanks{}


\date{\today}

\begin{abstract}
  We present a new method to observe direct experimental evidence of
  Jaynes--Cummings nonlinearities in a strongly dissipative cavity
  quantum electrodynamics system, where large losses compete with the
  strong light-matter interaction. This is a highly topical problem,
  particularly for quantum dots in microcavities where transitions
  from higher rungs of the Jaynes--Cummings ladder remain to be
  evidenced explicitly. We compare coherent and incoherent excitations
  of the system and find that resonant excitation of the detuned
  emitter make it possible to unambiguously evidence few photon
  quantum nonlinearities in currently available experimental systems.
\end{abstract}

\pacs{}
%\keywords{}

\maketitle

% to quote?:

% \cite{hennessy07a}
% \cite{yao09a}
% \cite{carmichael08a}

After reaching the strong coupling regime of the light-matter
interaction for optically active quantum dots (QDs) in high quality
microcavities~\cite{reithmaier04a,yoshie04a,peter05a} and
demonstrating its single-photon character~\cite{hennessy07a,press07a},
a major remaining challenge is to obtain clear and direct evidence of
quantum nonlinearities. Indirect manifestations have already been
provided in the form of photon
blockade~\cite{faraon08a,faraon10a,englund10a} or broadening of the
Rabi doublet due to excited states~\cite{kasprzak10a} but these are
unspecific with regard to their origin and whether such quantum
effects are described by the Jaynes--Cummings (JC) Hamiltonian,
\begin{equation}
  \label{eq1}
  H=\omega_a\ud{a}a+(\omega_a-\Delta)\ud{\sigma}\sigma
  +g(\ud{a}\sigma+a\ud{\sigma})\,,
\end{equation}
the paradigm of quantum interaction between quanta of light (with Bose
operator~$a$) and a two-level system ($\sigma$). Such quantum
nonlinearities have been demonstrated in cavity QED systems, such as
atoms~\cite{brune96a,schuster08a} or, perhaps most spectacularly, for
superconducting qubits~\cite{fink08a,bishop09a}, where the
fingerprints of JC physics have been observed, in particular the
anharmonic splitting of states with the number of
excitation~\cite{carmichael08a}. Only very recently have semiconductor
systems begun to exhibit this rich phenomenology~\cite{voltzPC}. The
difficulty of directly observing transitions between the different
rungs of the JC ladder in semiconductors can, presumably, be traced to
the strong dephasing in these systems. Indeed in the spectral domain
the uncertainty due to the short photon lifetime washes out completely
the weak square root dependence of the splitting of the excited
states. Any traces of the quantum interaction, lost in the energy of
the emitted photons, is however recovered in the statistics of the
emitted photons~\cite{schneebeli08a}.

In this Letter, we highlight that, although it is difficult to obtain
clear evidence of the JC ladder in state-of-the-art semiconductor
samples when detecting luminescence under incoherent excitation, it is
however possible to observe clear signatures of the higher rungs by
performing photon counting measurements to probe the statistics of the
emitted photons from the cavity.  We show that this process is optimum
when coherently exciting the detuned QD, which result in strong photon
bunching at the resonances of the JC ladder.  Our results provide a
route for experimentalists to test the suitability of JC model to
describe QD-cavity systems, which have displayed many variations from
their counterparts in atomic or superconducting cavity
QED~\cite{ota09b,winger09a}.

% Tailoring the interaction of light with matter at the single quantum
% level is one of the most pressing goal for both fundamental and
% applied research. A successful implementation sees a single quantum of
% excitation affecting the response of the system. Such an ultimate
% control is made possible in cavity Quantum Electrodynamics (cQED),
% where photons can be trapped in small numbers and brought to interact
% with a few quanta of excitation of an optical emitter. The latter has
% been realized with Rydberg atoms, semiconductor Quantum Dots (QD) and
% superconducting qubits, among others.

\begin{figure}[b]
  \centering
  \includegraphics[width=0.9\linewidth]{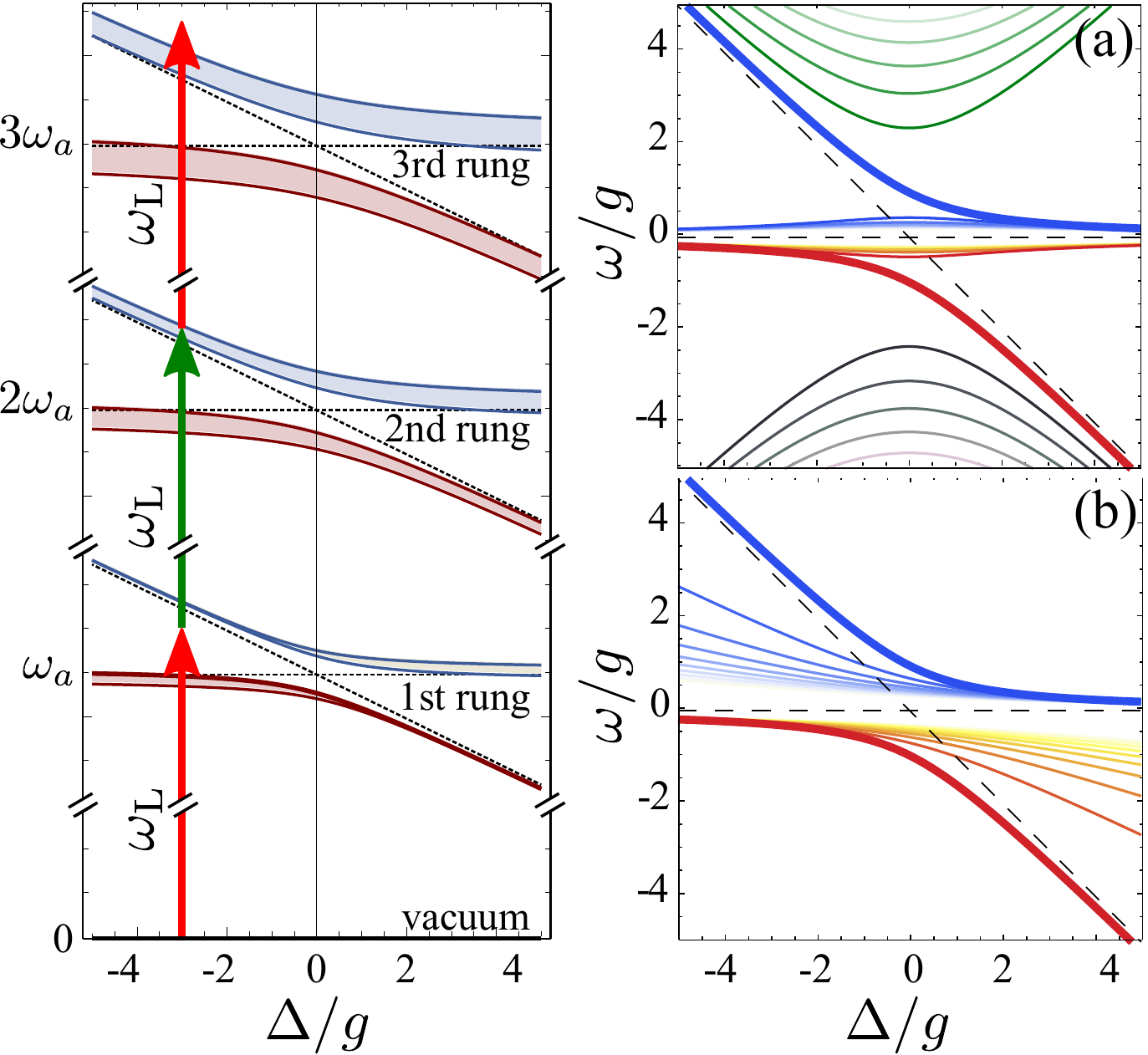}
  \caption{(color online) Energies and linewidths of the first three
    rungs of the dissipative JC ladder ($\gamma_a=g$) as function of
    detuning. The configuration of the two-photon blockade is
    indicated by the arrows.  Transitions energies and resonances of
    the JC ladder as a function of detuning, probed under incoherent
    (a) and coherent (b) excitation, respectively.  The thick solid
    lines are the upper and lower polariton lines of the first
    rung. The thin dotted lines are the bare (undressed) QD and
    cavity. They sandwich \emph{inner lines} and are encompassed by
    \emph{outer lines} from transitions higher in the ladder.}
  \label{fig1}
\end{figure}

The effect of dissipation can be introduced into the JC Hamiltonian
with Liouville equation $\partial_t\rho=\mathcal{L}(\rho)$, where the
so-called Liouvillian $\mathcal{L}$ adds a non-unitary evolution of
the density matrix $\rho$ to the Hamiltonian dynamics:
\small
\begin{equation}
  \label{eq2}
  \mathcal{L}(\rho)=i[\rho,H]
  +\frac{\gamma_a}{2}\mathcal{L}_a(\rho)+\frac{\gamma_\sigma}{2}\mathcal{L}_\sigma(\rho)
  +\frac{P_a}{2}\mathcal{L}_{\ud{a}}(\rho)+\frac{P_\sigma}{2}\mathcal{L}_{\ud{\sigma}}(\rho)
  \,,
\end{equation}
\normalsize
where~$\mathcal{L}_c(\rho)=2c\rho\ud{c}-\ud{c}c\rho-\rho
\ud{c}c$. This describes the decay (at rate $\gamma_a$ for the cavity
photon and $\gamma_\sigma$ for the QD exciton) or excitation ($P_a$
and $P_\sigma$) due to incoherent pumping~\cite{laussy08a}. Dephasing
can be included in this formalism with additional terms
$\mathcal{L}_{\ud{\sigma}\sigma}$ for pure
dephasing~\cite{laucht09b,auffeves09a,peter05a} or
$\mathcal{L}_{\ud{\sigma}a}$ for phonon induced
dephasing~\cite{arXiv_majumdar10a}. We have checked that unless these
quantities have very large values, they do not affect qualitatively
our findings.

As a result of finite lifetime, the energies of the JC system become
complex. They are obtained by diagonalizing the
Liouvillian~(\ref{eq2}):
\small
\begin{multline}
  \label{eq3}
  E_{\pm}^k=k\omega_a-\frac{\Delta}2-i\frac{(2k-1)\gamma_a+\gamma_\sigma}{4}\\
  \pm\sqrt{(\sqrt{k}g)^2-\left(\frac{\gamma_a-\gamma_\sigma}4+i\frac{\Delta}2\right)^2}\,,
\end{multline}
\normalsize
where $E_{\pm}^k$ corresponds to the $k$th rung of the system. This is
a generalization of the usual expression that neglects lifetime. It
shows that detuning behaves as an effective dissipation and, thus,
strong coupling is optimum at resonance. Following Eq.~(\ref{eq3}) we
plot the eigenenergies of a dissipative JC system ($\gamma_a=g$) in
Fig.~\ref{fig1} as function of detuning. Here, the broadening $\pm
2\Im(E_{\pm})$ is visualized as the width of the line. The first rung
($k=1$) is the familiar anticrossing of two coupled modes, that
describes equally well the linear quantum regime and a classical
system~\cite{zhu90a}. Higher rungs ($k>1$) reproduce the same pattern
with two variations with respect to the linear case (at resonance):
the splitting increases as $\sqrt{k}$ and the broadedning as $k
\gamma_a/g$. Since the increase of the coupling rate with the number
of excitations~$(k)$ is slower than the decoherence, climbing the
ladder makes it increasingly difficult to observe the quantum
features. One could circumvent this problem by decreasing $\gamma_a$,
but in typical semiconductor systems $\gamma_a \approx g
\gg\gamma_\sigma$, which is the configuration we will focus on in the
following (we shall in fact assume $\gamma_\sigma=0$). However, it is
advantageous if $\gamma_a$ is not too small: it increases signal
intensity and better preserves the statistics of the state prepared
inside the cavity. These two qualities are essential for working
quantum devices.
%Note that in
%Fig.~\ref{fig1}, although each rung is plotted
%exactly, according to Eq.~(\ref{eq3}) with
%broadening taken as the width of the line, $\pm\Im(E_{\pm})$, the
%spacing between two consecutive rungs has been shrunk for comfort of
%visualisation, and is in reality two or three orders of magnitude
%larger than shown (the case where the energy between two manifolds
%becomes comparable with the polariton splitting exists but belongs to
%an altogether different regime, called ``ultrastrong
%coupling''~\cite{ciuti06a, casanova10a}).
%
\begin{figure}[t]
  \centering
  \includegraphics[width=0.8\linewidth]{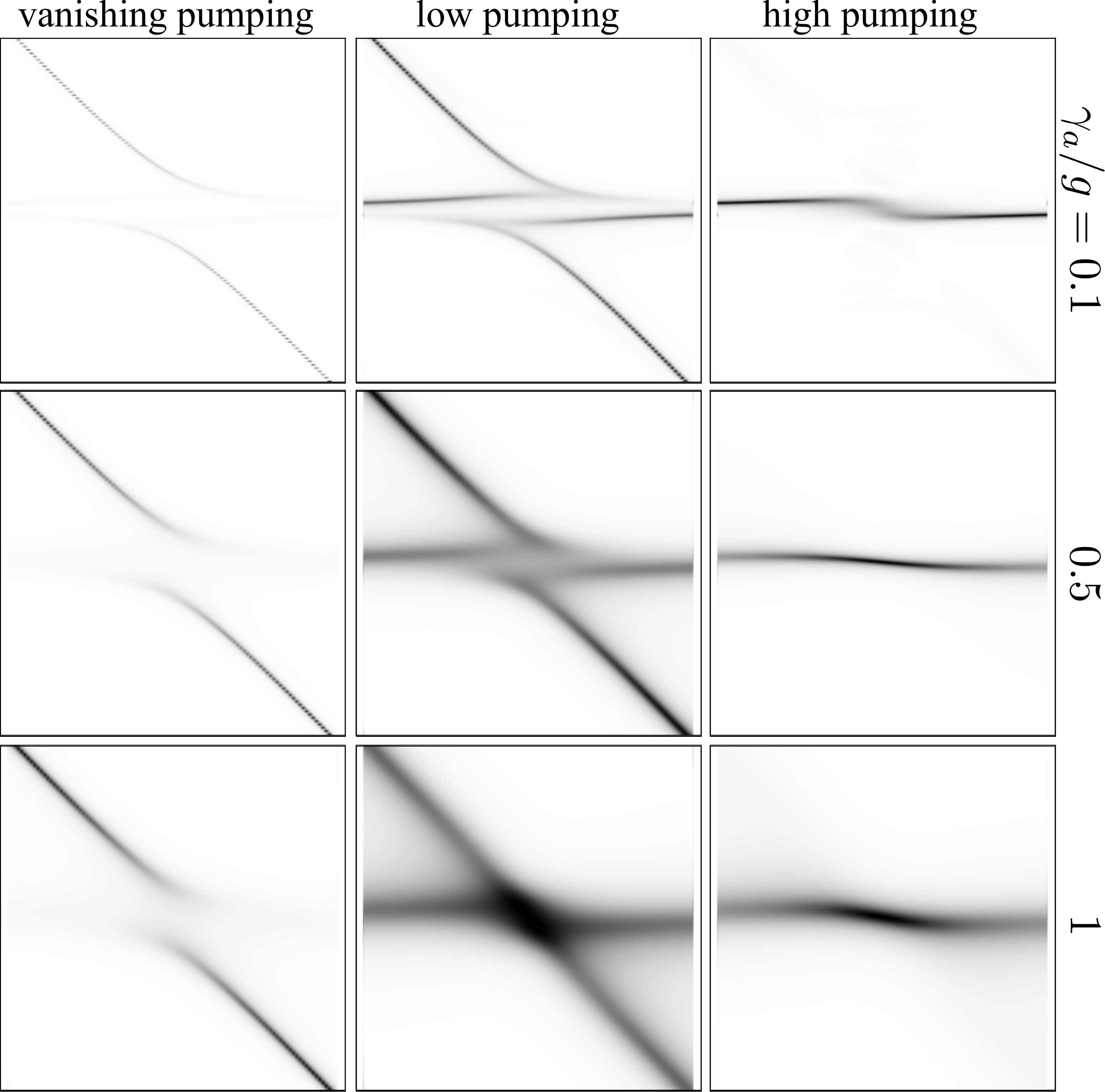}
  \caption{Incoherent excitation: cavity photoluminescence spectra for
    increasingly dissipative systems from upper to lower row
    ($\gamma_a/g=0.1$, 0.5 and 1), and for increasing excitation power
    from left to right column. Quantum nonlinearities are more clearly
    observed for a small, but nonvanishing incoherent
    excitation. Higher pumping brings the system into lasing or
    collapses the Rabi doublet. Only in very-strongly coupled systems
    does the photoluminescence reconstruct the JC ladder, albeit with
    the outer transitions being much suppressed in the cavity
    emission. Axes are not shown for clarity but are the same as
    Fig.~\ref{fig1}(a).}
  \label{fig2}
\end{figure}
Experimentally, the energy structure of the ladder cannot be observed
directly and the way it manifests itself depends on the kind of
measurement performed and how the system is excited. There are
countless variations of experiments, but most can be categorized in
\emph{incoherent} and \emph{coherent} excitation. In the former case,
one populates the energy levels of the system through relaxation of
charge carriers, and observes the emission at energies corresponding
to transitions between consecutive manifolds.  In the latter case, a
well defined energy is incident on the system and one observes its
direct response via a number of observables. The expected resonances
of the system in these two configurations are displayed in
Figs.~\ref{fig1}(a-b), respectively. The actual observation is a
combination of these lines, depending on the interplay of their
oscillator strengths and the fluctuations in population. Both have in
common the upper polariton (UP) and lower polariton (LP) lines (thick
lines) of the first rung, which are most easily excited and detected.
To prove the quantum character of this system, the field quantization
needs to be demonstrated, and this requires observing at least some of
the lines that arise from higher rungs of the ladder.

In the incoherent excitation case [Fig.~\ref{fig1}(a)], there are two
sets of additional lines, one in-between the Rabi doublet, the other
sandwiching it. The \emph{inner lines} are narrowly packed together
and are broader, since they arise from higher rungs, thus demanding an
extraordinarily good system to resolve them. The \emph{outer lines}
have a much higher splitting between them. However, the strength of
these transitions is strongly suppressed in the cavity emission, since
the emitter is de-excited at the same time as the photon is emitted,
whilst the cavity favours the sole emission of a photon.
%To climb the ladder in a
%dissipative system, a rather strong excitation is required, and this
%can strongly renormalise the energy levels, broaden them or, on the
%opposite, narrow them as a result of Bose-stimulation and onset of
%lasing. 
In Fig.~\ref{fig2}, we show the cavity photoluminescence spectra
$\langle\ud{a}(\omega)a(\omega)\rangle$~\cite{delvalle09a} as function
of detuning for increasing dissipation from top to bottom and for
various intensities of incoherent excitation from left to right. The
quality of the strong coupling ranges from significantly better than
is currently available ($\gamma_a/g=0.1$), via state-of-the-art
systems ($\gamma_a/g=0.5$~\cite{nomura10a}) to the typical value
available in many laboratories worldwide
($\gamma_a/g=1$~\cite{laucht09b}). These density plots show how the JC
structure of Fig.~\ref{fig1}(a) manifests itself in
photoluminescence. In all cases the outer lines are indeed
suppressed. The main features are the upper and lower polaritons. When
one tries to climb the ladder by increasing pumping, only in the very
best system can additional lines of the second rung be clearly
resolved.  In the case of $\gamma_a/g=0.5$, although a strong
deviation from the anticrossing is observed, no clear fingerprints of
the JC features are observed. At resonance, only a doublet is observed
(qualitatively similar to the Rabi doublet) and out of resonance, a
triplet is observed~\cite{arXiv_laussy11a}. This might in fact be
consistent with the experimental situation of Ref.~\cite{ota09b}, that
did not make any claim in this direction. For smaller strong-coupling,
although still deviating from crossing or anticrossing, there is again
no useful characterisation of the JC physics. At pumping levels higher
than those presented in Fig.~\ref{fig2}, the system moves into the
lasing regime~\cite{nomura10a} and the JC description is not adequate
anymore~\cite{delvalle10d}.

In the coherent excitation case, Fig.~\ref{fig1}(b), there are only
inner lines, but they are clearly separated from one another in the
energy-detuning space. These arise from multiple-photon
excitations. When the laser is at an energy~$\omega_\mathrm{L}$,
smaller than the upper polariton~$\omega_\mathrm{UP}$ energy, as
indicated by the arrows in Fig.~\ref{fig1}, it cannot excite the
system with one photon. However, if $2\omega =\Re(E_+^2)$, then it can
access the second rung by a two-photon excitation process.
%The separation of the lines is larger at larger detunings, and
%broadening of the exciton-like polariton is also smaller than at
%resonance, where it is half cavity-photon. However, the coupling is
%maximum at resonance. There is a tradeoff between detuning that
%provides the sought qualities to evidence quantum
%nonlinearity: small dephasing of the state and good coupling. 
%
\begin{figure}[t!]
  \centering
  \includegraphics[width=0.8\linewidth]{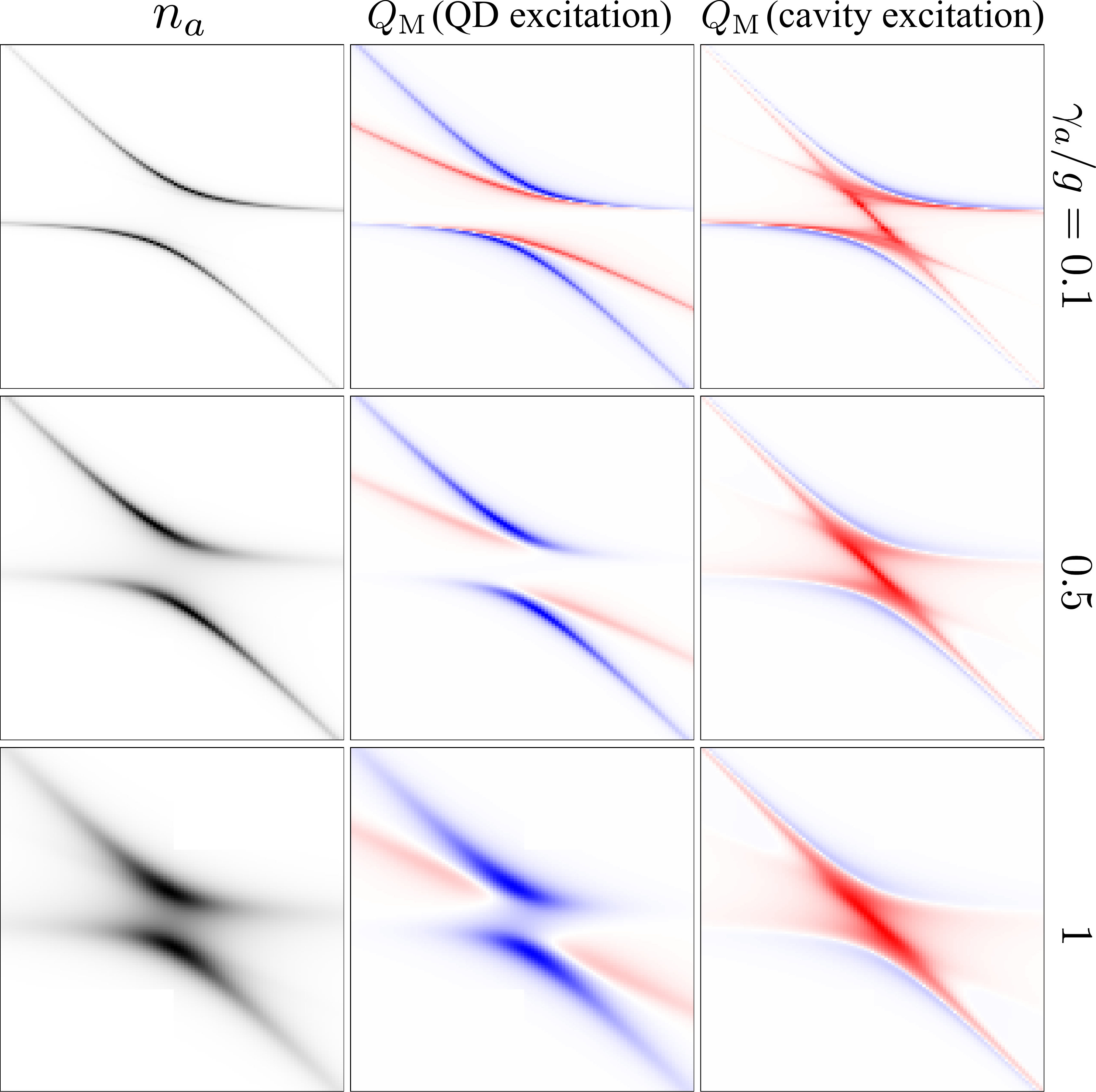}
  \caption{(color online) Coherent excitation: cavity intensity $n_a$
    (first column), and Mandel factor $Q_\mathrm{M}$ under coherent QD
    (second column) and cavity (third column) excitation, for
    increasingly dissipative systems. Blue and red refer to negative
    (antibunching) and positive (bunching) values, respectively. While
    intensity (or other observables such as absorption or scattering)
    elicit a response only from the bottom of the ladder, the photon
    statistics displays strong features from the 2nd rung when
    exciting the QD detuned from the cavity. The signature is then
    unambiguous even for very dissipative systems. For cavity
    excitation, bare modes dominate. Axes are not shown for clarity
    but are the same as Fig.~\ref{fig1}(b).}
  \label{fig3}
\end{figure}
At the point highlighted, the laser is blocked at the first rung, is
resonant with the second rung, and is blocked again at the third rung,
even when taking into account the large broadening of higher excited
manifolds.  This configuration, therefore, efficiently filters out the
two-photon fluctuation of the laser and performs a type of
\emph{two-photon blockade}, in analogy with the photon blockade
effect~\cite{imamoglu97a, birnbaum05a}, where blocking from the second
rung is used to produce a single-photon
source~\cite{faraon08a,faraon10a}. This scheme does not work so well
at resonance because of overlap of the transitions broadened by
dissipation.

\begin{figure}[t]
  \centering
  \includegraphics[width=\linewidth]{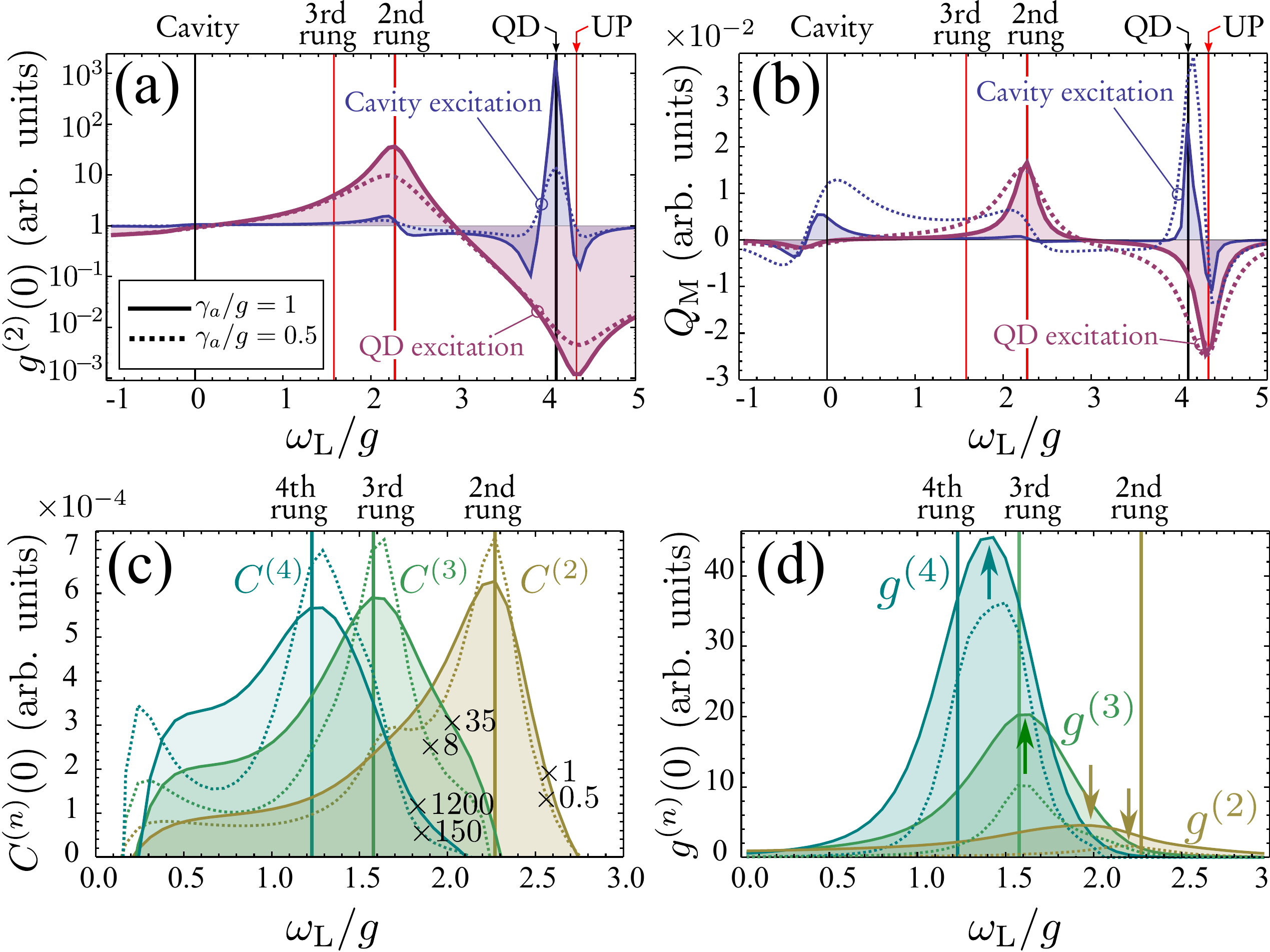}
  \caption{(color online) Quantum statistics of the cavity photons
    under coherent excitation for a good (dashed lines) and a typical
    (solid lines) system. (a) and (b): $g^{(2)}$ and $Q_\mathrm{M}$
    parameter, showing how the 2nd rung transition is more clearly
    identified under QD excitation. $Q_\mathrm{M}$ is preferable to
    $g^{(2)}$ as it takes into account the available signal. (c)
    Differential correlations~$C^{(n)}$ that peak sharply at the $n$th
    resonance, thus clearly identifying the higher states of the
    ladder. (d) $n$th order correlation functions $g^{(n)}$, that are
    loosely connected to the theoretical transitions. Maxima are
    indicated by arrows.}
  \vskip-.8cm
  \label{fig4}
\end{figure}

Furthermore, as for incoherent excitation, one should try to avoid
overly strong pumping for coherent excitation.  In the Hamiltonian it
is included by adding the term $\Omega_a\exp(i\omega
t)\ud{a}+\mathrm{h.c.}$ for coherent cavity
excitation~\cite{faraon08a,faraon10a,englund10a}. One can also excite
the QD coherently $\Omega_\sigma\exp(i\omega
t)\ud{\sigma}+\mathrm{h.c.}$, e.g., by side emission in a pillar
microcavity. At low driving intensity, one only sees clearly the lower
and upper polariton of the first rung, but for increasing pumping the
coherent excitation quickly dresses the states and distorts
Eq.~(\ref{eq3}).  In the first column of Fig.~\ref{fig3} we plot the
cavity population $n_a=\langle\ud{a}a\rangle$ when driving the system
coherently for increasingly dissipative systems from top to
bottom. The plots fail to reproduce the nonlinear features of
Fig.~\ref{fig1}(b), indicating that the intensity ($\propto
n_a$)---and in fact other observables involving first order
correlators such as reflectivity, transmission, absorption, etc.---are
not optimal to resolve the higher rungs of the JC ladder. In all these
cases, there is a strong response when the laser hits the polariton
resonances of the first rung, but otherwise the response is weak.
There is, however, a strong response in observables involving higher
order correlators (at zero time delay), of the type $G^{(n)}=\langle
a^{\dagger n}a^n\rangle$, which are linked to photon counting. One
usually deals with the normalized quantities,
$g^{(n)}=G^{(n)}/n_a^n$. In particular, $g^{(2)}$ is popular as the
standard to classify antibunched (non-classical) ($g^{(2)}=0$),
Poissonian (coherent) ($g^{(2)}=1$) and bunched (chaotic/thermal)
($g^{(2)}=2$) light sources.  Other related quantities are more useful
in this context~\cite{kubanek08a}, such as the Mandel
parameter~$Q_\mathrm{M}$:
\begin{equation}
  \label{eq5}
  Q_\mathrm{M}=(g^{(2)}-1)n_a
\end{equation}
which sign provides the (anti)bunched character of the statistics,
while also taking into account the available signal. The
$Q_\mathrm{M}$ parameter of the light emitted by the cavity is shown
in Fig.~\ref{fig3} for coherent excitation of the QD (middle column)
and of the cavity (right column). Remarkably, the middle column shows
that this measurement unravels the second rung of the JC ladder: when
the coherent excitation source is tuned to the 1st or the 2nd
resonance, the photon statistics sharply responds. As detailed in
Fig.~\ref{fig4}(a-b), the statistics changes its character from
antibunching on the polariton to bunching on the 2nd rung, caused by
the two-photon blockade configuration displayed in Fig.~\ref{fig1}. As
a result of the separation of the resonances, one can clearly observe
both resonances in this measurement with QD excitation.  The scheme is
much less efficient when exciting the cavity, with a strong response
on the bare QD, as can be seen both on the density plot in
Fig.~\ref{fig3} (right column) or on the cut in
Fig.~\ref{fig4}(a-b). This is due to a sudden drop of the intensity
when exciting the cavity at the energy of the QD.

These results show that even in very dissipative systems, where the
coupling rate is of the order of the decay rate, the 2nd rung can be
unambiguously observed in experiment, by combining detuning, QD
excitation and photon counting. Remarkably, this scheme can be
extended to higher rungs of the ladder, still with systems that
operate in the presence of significant dissipation. Measuring the
differential correlation function:
\begin{equation}
  \label{eq6}
  C^{(n)}=\langle a^{\dagger n}a^n\rangle-\langle a^{\dagger}a\rangle^n\,,
\end{equation}
which quantifies the deviation of coincidences from uncorrelated,
Poissonian events, one finds sharp resonances at the $n$th photon
resonance condition. In Fig.~\ref{fig4}(c) we plot $C^{(n)}$ up to the
fourth rung, still for coherent excitation of the dot, when it is
detuned by $\Delta/g=4$ from the cavity. Such high-order coincidences
can be measured by recently developed experimental techniques such as
photon counting using a streak camera~\cite{wiersig09a}. The signal is
increasingly difficult to obtain for higher order as it requires to
accumulate statistics for increasingly unlikely events (curves have
been rescaled as indicated in the figure). However, given a
sufficiently strong signal, one obtains sharp resonances precisely
located at the JC multi-photon resonances, even for very dissipative
systems such as those currently available.  The scheme is robust to
increased pumping, which broadens the resonances but does not
appreciably shift their maxima, and provides more signal. Although the
experiment to perform is the same, the quantity to analyse is
$C^{(n)}$ rather than $g^{(n)}$ which is shown in
Fig.~\ref{fig4}(d). One sees that $g^{(n)}$ increases with $n$, which
marks the higher quantumness of light emitted when hitting the higher
rungs. However, the signal also becomes exponentially dimmer as a
result.  Because of fluctuations to all orders, the resonances are
also not exactly mapped with the transitions [compare the agreement of
$g^{(2)}$ with the theoretical line at low pumping in (a) and its
disagreement at high pumping in (d)], in contrast to $C^{(n)}$ that
follow them accurately (except very close to resonance).

%

% To circumvent this, Kasprzak \emph{et al.} have recently used a
% four-wave mixing setup~\cite{kasprzak10a}. This allows to probe a
% second-order response of the Jaynes--Cummings system, which amounts to
% creating an initial state in the second rung. Time delay between
% pulses and filtering in time of the polarization allows to
% discriminate the dynamics of the higher rungs. We do not find however
% this technique to be at an advantage in the very dissipative system
% under consideration, as we discuss elsewhere~\cite{arXiv_schrapp11a}.

In conclusion, we have proposed an experimental scheme able to unravel
the JC nonlinearities in dissipative QD-cavity systems similar to
those currently available. We show that by analysing the photon
counting statistics of the light emitted by a cavity when the quantum
emitter is detuned and excited coherently, one can observe a clear and
unambiguous fingerprint of quantum nonlinearities. The demonstration
of JC physics of QDs in microcavities, will pave the way for
practical, working quantum information devices in the solid state.

We thank D.~Sanvitto for fruitful discussions, DFG via SFB-631,
Nanosystems Initiative Munich \& EU FP-7 via SOLID and Marie Curie
Initiative `SQOD'.

\bibliographystyle{apsrev}
\bibliography{Sci,books,arXiv,dJC}

\begin{thebibliography}{31}
\expandafter\ifx\csname natexlab\endcsname\relax\def\natexlab#1{#1}\fi
\expandafter\ifx\csname bibnamefont\endcsname\relax
  \def\bibnamefont#1{#1}\fi
\expandafter\ifx\csname bibfnamefont\endcsname\relax
  \def\bibfnamefont#1{#1}\fi
\expandafter\ifx\csname citenamefont\endcsname\relax
  \def\citenamefont#1{#1}\fi
\expandafter\ifx\csname url\endcsname\relax
  \def\url#1{\texttt{#1}}\fi
\expandafter\ifx\csname urlprefix\endcsname\relax\def\urlprefix{URL }\fi
\providecommand{\bibinfo}[2]{#2}
\providecommand{\eprint}[2][]{\url{#2}}

\bibitem[{\citenamefont{Reithmaier et~al.}(2004)\citenamefont{Reithmaier, Sek,
  L\"offler, Hofmann, Kuhn, Reitzenstein, Keldysh, Kulakovskii, Reinecker, and
  Forchel}}]{reithmaier04a}
\bibinfo{author}{\bibfnamefont{J.~P.} \bibnamefont{Reithmaier}}, et~al.,
  \bibinfo{journal}{Nature} \textbf{\bibinfo{volume}{432}},
  \bibinfo{pages}{197} (\bibinfo{year}{2004}).

\bibitem[{\citenamefont{Yoshie et~al.}(2004)\citenamefont{Yoshie, Scherer,
  Heindrickson, Khitrova, Gibbs, Rupper, Ell, Shchekin, and Deppe}}]{yoshie04a}
\bibinfo{author}{\bibfnamefont{T.}~\bibnamefont{Yoshie}}, et~al.,
  \bibinfo{journal}{Nature} \textbf{\bibinfo{volume}{432}},
  \bibinfo{pages}{200} (\bibinfo{year}{2004}).

\bibitem[{\citenamefont{Peter et~al.}(2005)\citenamefont{Peter, Senellart,
  Martrou, Lema\^itre, Hours, G\'erard, and Bloch}}]{peter05a}
\bibinfo{author}{\bibfnamefont{E.}~\bibnamefont{Peter}}, et~al.,
  \bibinfo{journal}{Phys. Rev. Lett.} \textbf{\bibinfo{volume}{95}},
  \bibinfo{pages}{067401} (\bibinfo{year}{2005}).

\bibitem[{\citenamefont{Hennessy et~al.}(2007)\citenamefont{Hennessy, Badolato,
  Winger, Gerace, Atature, Gulde, {F\u alt}, Hu, and {\u
  Imamo\=glu}}}]{hennessy07a}
\bibinfo{author}{\bibfnamefont{K.}~\bibnamefont{Hennessy}}, et~al.,
  \bibinfo{journal}{Nature} \textbf{\bibinfo{volume}{445}},
  \bibinfo{pages}{896} (\bibinfo{year}{2007}).

\bibitem[{\citenamefont{Press et~al.}(2007)\citenamefont{Press, G\"otzinger,
  Reitzenstein, Hofmann, L\"offler, Kamp, Forchel, and Yamamoto}}]{press07a}
\bibinfo{author}{\bibfnamefont{D.}~\bibnamefont{Press}}, et~al.,
  \bibinfo{journal}{Phys. Rev. Lett.} \textbf{\bibinfo{volume}{98}},
  \bibinfo{pages}{117402} (\bibinfo{year}{2007}).

\bibitem[{\citenamefont{Faraon et~al.}(2008)\citenamefont{Faraon, Fushman,
  Englund, Stoltz, Petroff, and Vuckovic}}]{faraon08a}
\bibinfo{author}{\bibfnamefont{A.}~\bibnamefont{Faraon}}, et~al.,
  \bibinfo{journal}{Nat. Phys.} \textbf{\bibinfo{volume}{4}},
  \bibinfo{pages}{859} (\bibinfo{year}{2008}).

\bibitem[{\citenamefont{Faraon et~al.}(2010)\citenamefont{Faraon, Majumdar, and
  Vuckovic}}]{faraon10a}
\bibinfo{author}{\bibfnamefont{A.}~\bibnamefont{Faraon}}, et~al.,
  \bibinfo{journal}{Phys. Rev. A} \textbf{\bibinfo{volume}{81}},
  \bibinfo{pages}{033838} (\bibinfo{year}{2010}).

\bibitem[{\citenamefont{Englund et~al.}(2010)\citenamefont{Englund, Majumdar,
  Faraon, Toishi, Stoltz, Petroff, and Vučković}}]{englund10a}
\bibinfo{author}{\bibfnamefont{D.}~\bibnamefont{Englund}}, et~al.,
  \bibinfo{journal}{Phys. Rev. Lett.} \textbf{\bibinfo{volume}{104}},
  \bibinfo{pages}{073904} (\bibinfo{year}{2010}).

\bibitem[{\citenamefont{Kasprzak et~al.}(2010)\citenamefont{Kasprzak,
  Reitzenstein, Muljarov, Kistner, Schneider, Strauss, H{\"o}fling, Forchel,
  and Langbein}}]{kasprzak10a}
\bibinfo{author}{\bibfnamefont{J.}~\bibnamefont{Kasprzak}}, et~al.,
  \bibinfo{journal}{Nat. Mater.} \textbf{\bibinfo{volume}{9}},
  \bibinfo{pages}{304} (\bibinfo{year}{2010}).

\bibitem[{\citenamefont{Brune et~al.}(1996)\citenamefont{Brune, Schmidt-Kaler,
  Maali, Dreyer, Hagley, Raimond, and Haroche}}]{brune96a}
\bibinfo{author}{\bibfnamefont{M.}~\bibnamefont{Brune}}, et~al.,
  \bibinfo{journal}{Phys. Rev. Lett.} \textbf{\bibinfo{volume}{76}},
  \bibinfo{pages}{1800} (\bibinfo{year}{1996}).

\bibitem[{\citenamefont{Schuster et~al.}(2008)\citenamefont{Schuster, Kubanek,
  Fuhrmanek, Puppe, Pinkse, Murr, and Rempe}}]{schuster08a}
\bibinfo{author}{\bibfnamefont{I.}~\bibnamefont{Schuster}}, et~al.,
  \bibinfo{journal}{Nat. Phys.} \textbf{\bibinfo{volume}{4}},
  \bibinfo{pages}{382} (\bibinfo{year}{2008}).

\bibitem[{\citenamefont{Fink et~al.}(2008)\citenamefont{Fink, G{\"o}ppl, Baur,
  Bianchetti, Leek, Blais, and Wallraff}}]{fink08a}
\bibinfo{author}{\bibfnamefont{J.~M.} \bibnamefont{Fink}}, et~al.,
  \bibinfo{journal}{Nature} \textbf{\bibinfo{volume}{454}},
  \bibinfo{pages}{315} (\bibinfo{year}{2008}).

\bibitem[{\citenamefont{Bishop et~al.}(2009)\citenamefont{Bishop, Chow, Koch,
  Houck, Devoret, Thuneberg, Girvin, and Schoelkopf}}]{bishop09a}
\bibinfo{author}{\bibfnamefont{L.~S.} \bibnamefont{Bishop}}, et~al.,
  \bibinfo{journal}{Nat. Phys.} \textbf{\bibinfo{volume}{5}},
  \bibinfo{pages}{105} (\bibinfo{year}{2009}).

\bibitem[{\citenamefont{Carmichael}(2008)}]{carmichael08a}
\bibinfo{author}{\bibfnamefont{H.}~\bibnamefont{Carmichael}},
  \bibinfo{journal}{Nat. Phys.} \textbf{\bibinfo{volume}{4}},
  \bibinfo{pages}{346} (\bibinfo{year}{2008}).

\bibitem[{\citenamefont{Volz and \Imamoglu}()}]{voltzPC}
\bibinfo{author}{\bibfnamefont{T.}~\bibnamefont{Volz}} and \bibnamefont{A}.~\bibnamefont{\Imamoglu},
  \bibinfo{howpublished}{private communication}.

\bibitem[{\citenamefont{Schneebeli et~al.}(2008)\citenamefont{Schneebeli, Kira,
  and Koch}}]{schneebeli08a}
\bibinfo{author}{\bibfnamefont{L.}~\bibnamefont{Schneebeli}}, et~al.,
  \bibinfo{journal}{Phys. Rev. Lett.} \textbf{\bibinfo{volume}{101}},
  \bibinfo{pages}{097401} (\bibinfo{year}{2008}).

\bibitem[{\citenamefont{Ota et~al.}(2009)\citenamefont{Ota, Kumagai, Ohkouchi,
  Shirane, Nomura, Ishida, Iwamoto, Yorozu, and Arakawa}}]{ota09b}
\bibinfo{author}{\bibfnamefont{Y.}~\bibnamefont{Ota}}, et~al.,
  \bibinfo{journal}{Appl. Phys. Express} \textbf{\bibinfo{volume}{2}},
  \bibinfo{pages}{122301} (\bibinfo{year}{2009}).

\bibitem[{\citenamefont{Winger et~al.}(2009)\citenamefont{Winger, Volz, Tarel,
  Portolan, Badolato, Hennessy, Hu, Beveratos, Finley, Savona
  et~al.}}]{winger09a}
\bibinfo{author}{\bibfnamefont{M.}~\bibnamefont{Winger}}, et~al.,
  \bibinfo{journal}{Phys. Rev. Lett.} \textbf{\bibinfo{volume}{103}},
  \bibinfo{pages}{207403} (\bibinfo{year}{2009}).

\bibitem[{\citenamefont{Laussy et~al.}(2008)\citenamefont{Laussy, del Valle,
  and Tejedor}}]{laussy08a}
\bibinfo{author}{\bibfnamefont{F.~P.} \bibnamefont{Laussy}}, et~al.,
  \bibinfo{journal}{Phys. Rev. Lett.} \textbf{\bibinfo{volume}{101}},
  \bibinfo{pages}{083601} (\bibinfo{year}{2008}).

\bibitem[{\citenamefont{Laucht et~al.}(2009)\citenamefont{Laucht, Hauke,
  Villas-B\^oas, Hofbauer, B\"ohm, Kaniber, and Finley}}]{laucht09b}
\bibinfo{author}{\bibfnamefont{A.}~\bibnamefont{Laucht}}, et~al.,
  \bibinfo{journal}{Phys. Rev. Lett.} \textbf{\bibinfo{volume}{103}},
  \bibinfo{pages}{087405} (\bibinfo{year}{2009}).

\bibitem[{\citenamefont{Auff\`eves et~al.}(2009)\citenamefont{Auff\`eves,
  G\'erard, and Poizat}}]{auffeves09a}
\bibinfo{author}{\bibfnamefont{A.}~\bibnamefont{Auff\`eves}}, et~al.,
  \bibinfo{journal}{Phys. Rev. A} \textbf{\bibinfo{volume}{79}},
  \bibinfo{pages}{053838} (\bibinfo{year}{2009}).

\bibitem[{\citenamefont{Majumdar et~al.}(2010)\citenamefont{Majumdar, Gong,
  Kim, and Vuckovic}}]{arXiv_majumdar10a}
\bibinfo{author}{\bibfnamefont{A.}~\bibnamefont{Majumdar}}, et~al.,
  \bibinfo{journal}{arXiv:1012.3125}  (\bibinfo{year}{2010}).

\bibitem[{\citenamefont{Zhu et~al.}(1990)\citenamefont{Zhu, Gauthier, Morin,
  Wu, Carmichael, and Mossberg}}]{zhu90a}
\bibinfo{author}{\bibfnamefont{Y.}~\bibnamefont{Zhu}}, et~al.,
  \bibinfo{journal}{Phys. Rev. Lett.} \textbf{\bibinfo{volume}{64}},
  \bibinfo{pages}{2499} (\bibinfo{year}{1990}).

\bibitem[{\citenamefont{del Valle et~al.}(2009)\citenamefont{del Valle, Laussy,
  and Tejedor}}]{delvalle09a}
\bibinfo{author}{\bibfnamefont{E.}~\bibnamefont{del Valle}}, et~al.,
  \bibinfo{journal}{Phys. Rev. B} \textbf{\bibinfo{volume}{79}},
  \bibinfo{pages}{235326} (\bibinfo{year}{2009}).

\bibitem[{\citenamefont{Nomura et~al.}(2010)\citenamefont{Nomura, Kumagai,
  Iwamoto, Ota, and Arakawa}}]{nomura10a}
\bibinfo{author}{\bibfnamefont{M.}~\bibnamefont{Nomura}}, et~al.,
  \bibinfo{journal}{Nat. Phys.} \textbf{\bibinfo{volume}{6}},
  \bibinfo{pages}{279} (\bibinfo{year}{2010}).

\bibitem[{\citenamefont{Laussy et~al.}(2011)\citenamefont{Laussy, Laucht, del
  Valle, Finley, and Villas-Bôas}}]{arXiv_laussy11a}
\bibinfo{author}{\bibfnamefont{F.}~\bibnamefont{Laussy}}, et~al.,
  \bibinfo{journal}{arXiv:1102.3874}  (\bibinfo{year}{2011}).

\bibitem[{\citenamefont{del Valle and Laussy}(2010)}]{delvalle10d}
\bibinfo{author}{\bibfnamefont{E.}~\bibnamefont{del Valle}} et~al.,
  \bibinfo{journal}{Phys. Rev. Lett.} \textbf{\bibinfo{volume}{105}},
  \bibinfo{pages}{233601} (\bibinfo{year}{2010}).

\bibitem[{\citenamefont{\Imamoglu et~al.}(1997)\citenamefont{\Imamoglu,
  Schmidt, Woods, and Deutsch}}]{imamoglu97a}
\bibinfo{author}{\bibfnamefont{A.}~\bibnamefont{\Imamoglu}}, et~al.,
  \bibinfo{journal}{Phys. Rev. Lett.} \textbf{\bibinfo{volume}{79}},
  \bibinfo{pages}{1467} (\bibinfo{year}{1997}).

\bibitem[{\citenamefont{Birnbaum et~al.}(2005)\citenamefont{Birnbaum, Boca,
  Miller, Boozer, Northup, and Kimble}}]{birnbaum05a}
\bibinfo{author}{\bibfnamefont{K.}~\bibnamefont{Birnbaum}}, et~al.,
  \bibinfo{journal}{Nature} \textbf{\bibinfo{volume}{436}}, \bibinfo{pages}{87}
  (\bibinfo{year}{2005}).

\bibitem[{\citenamefont{Kubanek et~al.}(2008)\citenamefont{Kubanek,
  Ourjoumtsev, Schuster, Koch, Pinkse, Murr, and Rempe}}]{kubanek08a}
\bibinfo{author}{\bibfnamefont{A.}~\bibnamefont{Kubanek}}, et~al.,
  \bibinfo{journal}{Phys. Rev. Lett.} \textbf{\bibinfo{volume}{101}},
  \bibinfo{pages}{203602} (\bibinfo{year}{2008}).

\bibitem[{\citenamefont{Wiersig et~al.}(2009)\citenamefont{Wiersig, Gies,
  Jahnke, A{\ss}mann, Berstermann, Bayer, Kistner, Reitzenstein, Schneider,
  H\"ofling et~al.}}]{wiersig09a}
\bibinfo{author}{\bibfnamefont{J.}~\bibnamefont{Wiersig}}, et~al.,
  \bibinfo{journal}{Nature} \textbf{\bibinfo{volume}{460}},
  \bibinfo{pages}{245} (\bibinfo{year}{2009}).

\end{thebibliography}
\end{document}